# Extended RBAC With Blob Storage On Cloud

Mamoon Rashid
Research Scholar
Department Of Computer Science Engineering
Ramgharia Institute of Engineering and Technology
Phagwara, Punjab, India.
Email: mamoon873@gmail.com.

Er. Rishma Chawla
Assistant Professor
Department Of Computer Science Engineering
Ramgharia Institute of Engineering and Technology
Phagwara, Punjab, India.
Email: chawlan02@yahoo.com.

**Abstract— Role-based access control (RBAC) models have generated a great interest in the security community as a powerful and generalized approach to security management and ability to model organizational structure and their capability to reduce administrative expenses. In this paper, we highlight the drawbacks of latest developed RBAC models in terms of access control and authorization and later provide a more viable extended-RBAC model, which enhances and extends its powers to make any system more secure by adding valuable constraints. Later the Blobs are stored on cloud server which is then accessed by the end users via this Extended RBAC model.**

*Keywords-Authorization,RBAC,Blobs,Server*

I INTRODUCTION

Authorization and access control has always been a fundamental security technique in systems in which multiple users share access to common resources. Authorization is the process of expressing security policies that determine whether a subject (e.g., process, computer, human user, etc.) is allowed to perform an operation (e.g., read, write, execute, delete, search, etc.) on an object (e.g., a tuple in a database, a table, a file, a service, and, more generally, any resource of the system). These policies define the subject's permissions (rights to carry out an operation on an object) in a computer system. Access control is the process of enforcing these policies in order to achieve the desired level of security. Managing and administering the users' privileges is one of the most challenging tasks in access control. Several access control models have been proposed, such as, discretionary and mandatory access control models (DAC and MAC), Clark-Wilson model, Lipner's Integrity model, Chinese wall model, Task based models, and Role Based Access Control models and RBAC has further been extended up to some level. Among these models Role-based access control (RBAC) models have been receiving attention as they provide systematic access control security through a proven and increasingly predominant technology for commercial organizations. One of the main advantages of the RBAC over other access control models is the ease of its security administrations. RBAC models are policy neutral [5]; they can support different authorization policies including mandatory and discretionary through the appropriate role configuration. In spite of the success of the RBAC, researchers have determined that there are still many application security requirements that are not addressed by the existing RBAC models [6]. In the past few years, several RBAC extensions have been proposed to address such security requirements [2, 3, 4, 6, 8, 13, and 14]. Although, these extensions geared and enhanced basic RBAC model, however we find some applications where individually these extended models fall short. We will uncover the loophole in existing extended RBAC model and later will add a mechanism to resolve such an issue.

II BASICS ABOUT ROLE BASED ACCESS CONTROL

Sandhu et al [1] proposed RBAC 96 which is a family of four constitutes models. In RBAC permissions are associated with roles (the intermediate concept of roles can be seen as collections of permissions), and users are made members of appropriate roles. The notion of role is an enterprise or organizational concept. The definition of role is quoted from Sandhu et al. [1]: A role is a job function or job title within the organization with some associated semantics regarding the authority and responsibility conferred on a member of the role. Permissions are not directly assigned to users; instead they are assigned to roles. RBAC comprise a family of four references models:

RBAC0: contains the core concepts of the Model. It is the minimum requirement for any system that exploits features of RBAC. Users (U), roles (R), and permissions (P) are three sets of entities and the relations between these entities are defined by User-Role Assignment and Permission-Role Assignment [1]. These sets and relations are the main concepts of the RBAC. A user can be member of many roles and each role can have many users. A user can invoke multiple sessions within a session a user can invoke set of roles but each session belongs to only one user. Permission can be assigned to many roles and a role can have many permissions.

RBAC1: adds to RBAC0 a role hierarchy (RH). Role hierarchies are an important concept for structuring roles to represent organization users responsibly and degree of authority.

RBAC2: introduces the concept of constraints. RBAC adds static (not related to sessions) and dynamic (related to sessions) constraints between core concepts [1] . These constraints are considered to be the principle motivation for RBAC because constraints are powerful mechanism to lay out higher-level organizational mechanism [1].Constraints can be applied to User-Role Assignment, Permission-Role Assignment and session.

RBAC3: includes all aspects of RBAC0, RBAC1 and RBAC2 and it is called a unified model of RBAC. RBAC3 combine RBAC1 and RBAC2 to combine both role hierarchy and constraints. In this model constraints can be applied to the role hierarchy in addition to the constraints in RBAC2.

E-RBAC: E-RBAC model was presented for RBAC 96 model and it filled the role authorization shortage in RBAC96. In E-RBAC users can also direct for authorization. For example, when the authority is authorized after a specific request to access the user's system resources, the system at the same time can judge whether the users own role or whether the user has authority access to the module's functions, as long as one is given the authority between the users and their role, to allow access.

## III BLOB STORAGE ON CLOUD SERVER

Blob storage is a service for storing large amounts of unstructured data that can be accessed from anywhere in the world via HTTP or HTTPS. A single blob can be hundreds of gigabytes in size, and a single storage account can contain up to 100TB of blobs. Common uses of Blob storage include:

- Serving images or documents directly to a browser
- Storing files for distributed access
- Streaming video and audio
- Performing secure backup and disaster recovery

The Blob service contains the following components:

**Storage Account:** All access to Cloud Storage is done through a storage account. This is the highest level of the namespace for accessing blobs. An account can contain an unlimited number of containers, as long as their total size is under 100TB.

**Container:** A container provides a grouping of a set of blobs. All blobs must be in a container. An account can contain an unlimited number of containers. A container can store an unlimited number of blobs.

**Blob:** A file of any type and size. There are two types of blobs that can be stored in Cloud Storage: block and page blobs. Most files are block blobs. A single block blob can be up to 200GB in size. Page blobs, another blob type, can be up to 1TB in size, and are more efficient when ranges of bytes in a file are modified frequently.

## IV PROBLEM FORMULATION

Although current RBAC models restrict a user to access the resources if it is not assigned as a member to any particular role of the architecture, however the RBAC nowhere defines that how many users could be assigned to each role. This limitation largely affects the architecture not only in terms of its security but also when one accesses the resources on a shared server where it affects the network bandwidth also. RBAC also allows users to access resources based on the roles. All users are made members of roles and permissions are also associated with roles. Later when any user wants to access any resource, its access depends on the constraints imposed on that particular role and it is imposed on all other users of that role also. This limitation makes architecture quite ordinary as there is no such mechanism to provide different level of accesses to different users under one particular role. Also RBAC does not impose any transaction limit for users to access available resources under specified roles. This limitation leads to insecurity of the system where resources can be accessed at free will even if login of any existing user is hacked or uncovered. Keeping all these flaws of RBAC into consideration, there is need of an architectural system which will address all these issues and hence will result as a more powerful RBAC model.

## V PROPOSED ERBAC MODEL

This Extended RBAC Model is proposed to address aforesaid problems so as to enhance its robustness. Firstly in this model during the creation of any new role, the constraint is imposed on role to decide how many users can access this role later. Here an organization can make use of this feature to provide membership of role to only such users which are beneficial for the organization. Secondly another constraint is imposed on all the normal users who can access resources later so that they can access the organizational resources on a limited basis for some specific time intervals. Although it looks that it will result in inconvenience to users as they can make limited transactions and thus cannot access the resources at free will. However it is one of the strong dimensions where we can make RBAC more powerful. This is because imposing transactions limits will reduce the chances of minimizing loss of resources of any particular user if cannot be saved and eliminated in totality. However here organization can assign the limit based on the usual usage of users. Next whenever any user is made member of any particular role, then at that time we are providing an option there to rate this user based on his membership status. This feature will largely help an organization to give different accesses to users irrespective of belonging to the same role. This decision is to be taken on the basis of premium membership to be attained by the user. Here by using this feature all normal users can access resources only to that extent to which that particular role will allow them. However the premium users will get access more than to role limit and will be decided on the basis of their premium membership. Finally in this proposed system we are planning to use the Blob service of resources so as use it service as on Cloud. Later we will impose all the added features on these blobs so as to check its integrity and robustness to evolve as an enhanced model. All these features are summed up in the following flow based diagram:

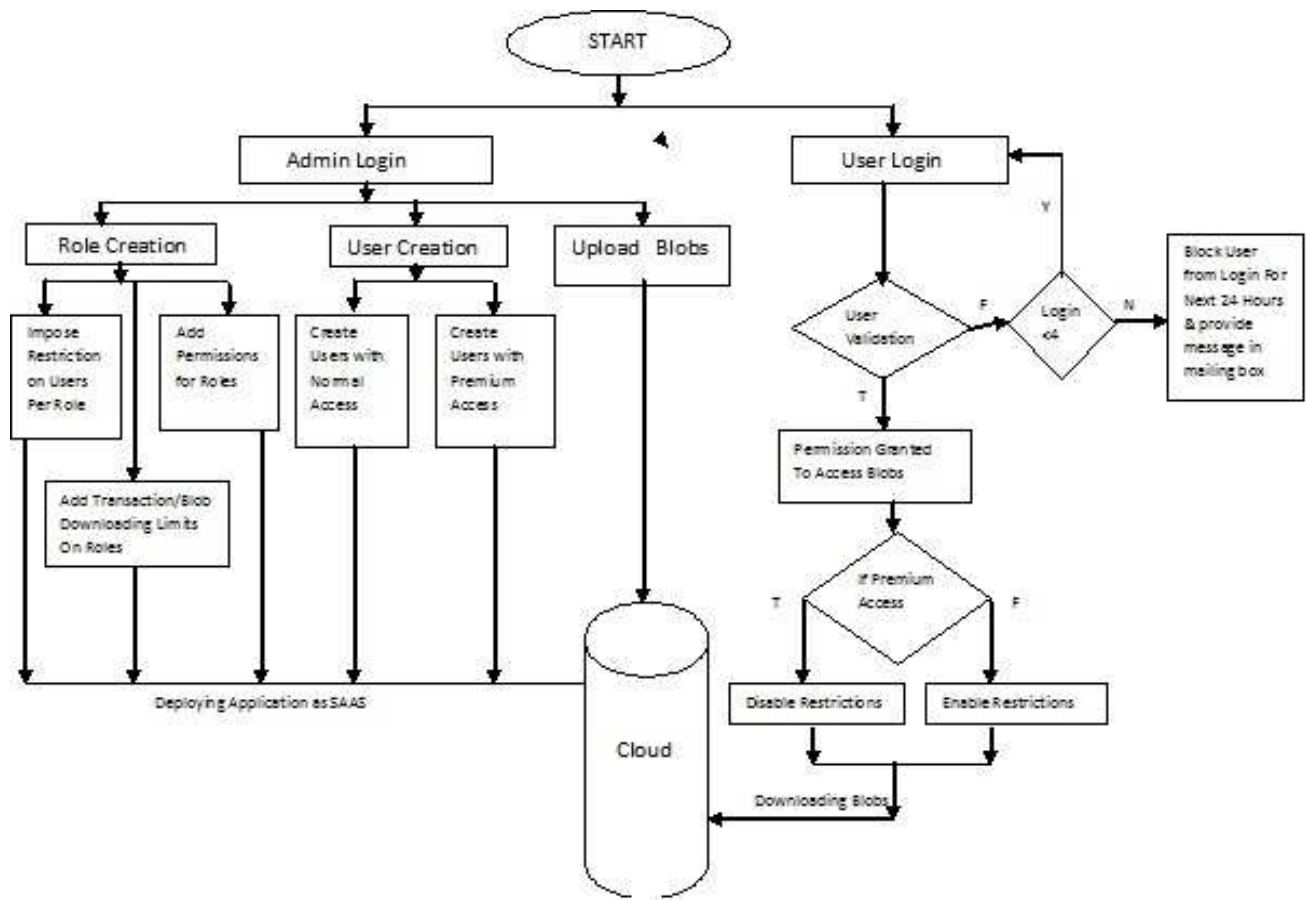

## VI CONCLUSION:

This proposed model has outlined a sketch for new RBAC which addresses the security features for any multi-centric application. We have investigated the state-of-art of the access control models. In particular, we have investigated the current RBAC extensions as they are most influential authorization models in the security community. This proposed model, however, showed that, all or most of the existing RBAC extensions are not suitable for specifying security requirements of that application. Although this study might not be exhaustive, however we believe that this model will provide another picture of the RBAC will result into much enhanced and more powerful model than other existing models. The investigations in this study open the doors to the software stake holders to identify and realize the added features which can make their organizational architecture more robust than existing RBAC models.

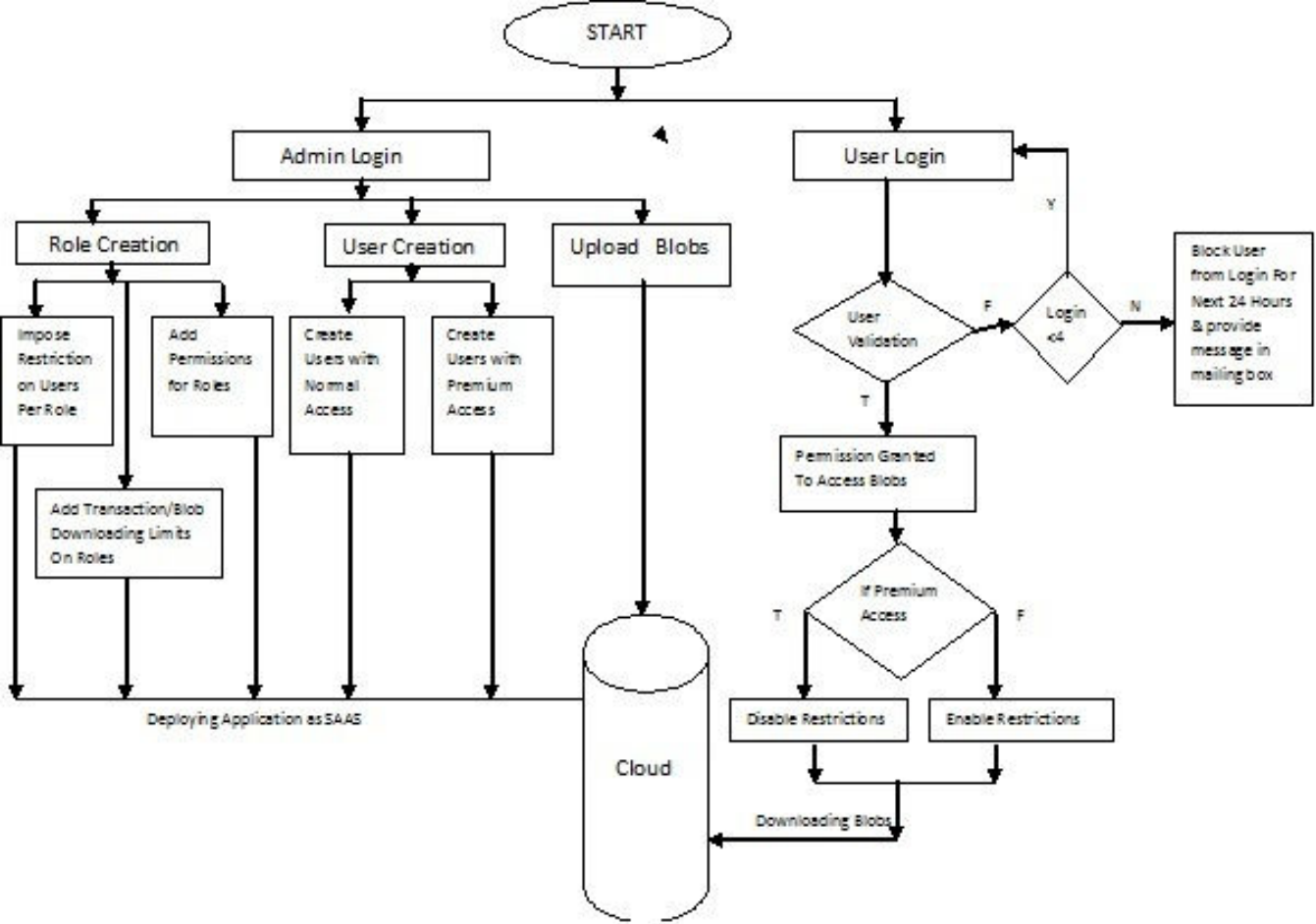

Figure 1